\documentstyle[preprint,prd,aps]{revtex}
\begin{document}
\title{A Supersymmetric SO(10) Model with Inflation and Cosmic Strings}
\author{Rachel Jeannerot\thanks{e-mail: R.Jeannerot@damtp.cam.ac.uk} \\
        {\normalsize{ Department of Applied Mathematics and
Theoretical Physics, Cambridge University,}}\\
        {\normalsize{ Silver Street, Cambridge, CB3 9EW, UK}}}
\date{\today}
\maketitle
\begin{abstract}
We have built a supersymmetric SO(10) model consistent with
cosmological observations. The model gives rise to a false vacuum hybrid
inflationary scenario which solves the monopole problem. We argue that this
type of inflationary scenario is generic in supersymmetric SO(10) model, and
arises naturally from the theory. Neither any external field nor any
external symmetry has to be added. It can just be a consequence of
the theory. In our specific model, at the end of inflation, cosmic
strings form. The properties of the strings are presented. The cosmic
background radiation anisotropies induced by the inflationary perturbations
and the cosmic strings are estimated. The model produces a stable
lightest superparticle and a very light left-handed neutrino which may
serve as the cold and hot dark matter. The properties of a mixed
cosmic string-inflationary large scale structure formation scenario
are discussed.
\end{abstract}

\section{introduction}

Supersymmetric SO(10) models have received much interest in the
past ten years. SO(10) is the minimal Grand Unified gauge group which
unifies all kinds of matter, thanks to its 16-dimensional spinorial
representation to which all fermions belonging to a single family can
be assigned. The running of the gauge coupling constants measured at
LEP in the Minimal Supersymmetric Standard Model with supersymmetry
broken at $10^3$ GeV merge in a single point at $2 \times 10^{16}$ GeV
\cite{meet}, hence strongly favoring supersymmetric versions of Grand
Unified Theories (GUTs). The doublet-triplet splitting can be easily achieved
in supersymmetric SO(10), through the Dimopoulos-Wilczek mechanism
\cite{DW}. The fermions masses can be beautifully derived
\cite{masses}. The gauge hierarchy problem can be solved
\cite{shafiun}. A $Z_2$ symmetry subgroup of the $Z_4$ center of
SO(10) can be left unbroken down to low energies, provided  only ``safe''
representations \cite{Martin} are used to implement the symmetry breaking
from SO(10) down to the standard model gauge group. We should point out
here that, when we write
SO(10), we really mean its universal covering group, spin(10).
The $Z_2$ symmetry can suppress rapid proton decay and provide a cold dark
matter
candidate, stabilizing the Lightest SuperParticle (LSP). Finally,
introducing a pair of Higgs in the $126 + \overline{126}$ representations
can give a superheavy Majorana mass to the right-handed neutrino, thus
providing a hot dark-matter candidate and solving the solar neutrino
problem through the Mikheyev-Smirnov-Wolfenstein (MSW) mechanism \cite{MSW}.
All these features make supersymmetric SO(10) models very
attractive.

In a recent paper \cite{paper}, we have constrained supersymmetric SO(10)
models using cosmological arguments. We have in particular studied the
formation of topological defects in all possible symmetry breaking
patterns from supersymmetric SO(10) down to the standard model,
considering no more than one intermediate symmetry breaking
scale.
Domain walls and monopoles lead to a cosmological catastrophe, while
cosmic strings can explain large scale structures, part of the baryon asymmetry
of the universe, and thermal fluctuations in the Cosmic
Background Radiation (CBR). Since SO(10) is simply connected and the
standard model gauge group involves an unbroken U(1) symmetry, which remains
unbroken down to low energy, all symmetry breaking patterns from
supersymmetric SO(10) down to the standard model automatically lead to
the formation of topologically stable monopoles via the Kibble
mechanism \cite{Kibble}. All supersymmetric SO(10) models are therefore
cosmologically irrelevant without invoking some mechanism for the
removal of the monopoles, such as an inflationary scenario.
The conclusions in \cite{paper}, is that there are only two possibilities
for breaking SO(10) down to the standard model which are consistent
with observations. SO(10) can break
via $SU(3)_c \times SU(2)_L \times SU(2)_R \times U(1)_{B-L}$, here SO(10)
must be broken with a combination of a 45-dimensional Higgs
representation and a 54-dimensional one, and via $SU(3)_c
\times SU(2)_L \times U(1)_R \times U(1)_{B-L}$. In these models,
the intermediate symmetry group must be broken down to the standard model
gauge group with unbroken matter parity, $SU(3)_c
\times SU(2)_L \times U(1)_Y \times Z_2$. In supergravity SO(10) models,
the breaking of SO(10) via flipped SU(5) is also possible.

In this paper,
we study a supersymmetric SO(10) model
involving an intermediate $SU(3)_c \times SU(2)_L \times U(1)_R \times
U(1)_{B-L}$ symmetry. The resultant cosmological model is
compatible with observations.

In section \ref{sec-inflation}, we describe a hybrid inflationary scenario
first introduced by \cite{L}, and we argue that this type of
inflationary scenario occurs naturally in global supersymmetric SO(10) models.
Neither any external field nor any external symmetry has to be imposed.
Inflation is driven by a scalar field singlet
under SO(10).

In the next sections, we construct a specific supersymmetric SO(10) model,
as mentioned above. The latter aims to be consistent with observations. In
section \ref{sec-topo}
we study the symmetry breaking pattern. We conclude on the proton lifetime
and on a hot dark-matter candidate
provided by the model. Using homotopy theory, we find topological defects
which form according to the Kibble mechanism \cite{Kibble}.

In section \ref{sec-building}, we explain how to implement the symmetry
breaking pattern which solves the doublet-triplet splitting and includes
the inflationary scenario described in section \ref{sec-inflation}. We
write down the superpotential and find its global minimum with
corresponding Higgs VEVs.

In section \ref{sec-infl}, we evaluate the dynamics of the symmetry
breaking and inflationary scenario, studying the scalar potential. It is shown
that the monopole
problem may be solved and that cosmic strings form at the end of inflation.

In section \ref{sec-strings}, we give general properties of the strings
formed at the end of inflation. In particular, we study the possibility
that the strings may be superconducting.

In section \ref{sec-observations}, we estimate the observational
consequences. The temperature fluctuations
in the CBR due to the mixed inflation-cosmic
strings scenario are evaluated.
Using the temperature fluctuations measured by COBE we find values
for the scalar coupling constant, the scale at which the strings
formed and the strings mass per unit length. We specify the dark-matter
present in the model and give a
qualitative discussion of the large-scale structure formation scenario
in this model.

We finally conclude in section \ref{sec-conclusion}.

In order to simplify the notation, we shall make use of the following
\begin{eqnarray}
a. \qquad & 4_c 2_L 2_R &\equiv SU(4)_c \times SU(2)_L \times SU(2)_R \nonumber
\\
b. \qquad & 3_c 2_L 2_R 1_{B-L} &\equiv SU(3)_c \times SU(2)_L \times SU(2)_R
\times U(1)_{B-L}
\nonumber \\
c. \qquad & 3_2 2_L 1_R 1_{B-L} &\equiv SU(3)_c \times SU(2)_L \times U(1)_R
\times U(1)_{B-L} \nonumber \\
d. \qquad & 3_c 2_L 1_Y Z_2 & \equiv SU(3)_c \times SU(2)_L \times U(1)_Y
\times Z_2 \nonumber \\
e. \qquad & 3_c 1_Q Z_2 & \equiv SU(3)_c \times U(1)_Q \times Z_2 \nonumber
\end{eqnarray}

\section{Inflation in supersymmetric SO(10) models}
\label{sec-inflation}

In this section, we argue that false vacuum hybrid inflation, with a
superpotential in the inflaton
sector similar to that studied in \cite{Ed,Shafi}, is a natural mechanism for
inflation in global supersymmetric SO(10) models. Neither any external
field nor any external symmetry has to be imposed, it can just be a
consequence of the theory.

The first thing to note in SO(10) models, is that the rank
of SO(10) is greater than one unit from the rank of the standard model
gauge group. The rank of SO(10) is five, whereas the rank of the standard
model gauge group $SU(3)_c \times SU(2)_L
\times U(1)_Y (\times Z_2)$ is 4. In other words, SO(10) has an additional U(1)
symmetry, named $U(1)_{B-L}$, compared to the standard model gauge group.
Therefore the rank of the
group must be lowered by one unit at some stage of the symmetry breaking
pattern, i.e. $U(1)_{B-L}$ must be broken.  This can be
done using a pair of $16 + \overline{16}$ Higgs
representations or by a pair of $126 + \overline{126}$
representations. If a $16 + \overline{16}$ pair
of Higgs fields are used, then the $Z_2$ symmetry, subgroup of both the $Z_4$
center of SO(10) and of $U(1)_{B-L}$, playing the role of matter parity, is
broken.
On the other hand, if a $126 + \overline{126}$ pair of Higgs fields are used,
 then the $Z_2$ symmetry can be kept unbroken down to low energy
if only safe representations \cite{Martin} are used to implement
the full symmetry breaking pattern, such as the 10, the 45, the 54 or the
210-dimensional representations. If a $126 + \overline{126}$ pair of
Higgs fields are used, the right-handed neutrino can get
a superheavy Majorana mass, and the solar neutrino problem can be solved
via the MSW mechanism \cite{MSW}.

In order to force the VEVs of the  $16 + \overline{16}$ or $126 +
\overline{126}$
pair of Higgs fields, needed to lower the rank of the group, to be the order of
the GUT scale, we can use a scalar field ${\cal S}$ singlet under SO(10). The
superpotential can be written as follows,
\begin{equation}
W_1 = \alpha {\cal S} \overline{\Phi} \Phi - \mu^2 {\cal S}
\label{eq:inflation}
\end{equation}
where $\Phi + \overline{\Phi}$ stand for a $16 + \overline{16}$  or
a $126 + \overline{126}$ pair of Higgs fields, and the field ${\cal S}$
is a scalar field singlet under SO(10). The constants $\alpha$ and $\mu$
are assumed to be both positive and must satisfy ${\mu \over \sqrt{\alpha}}
\sim (10^{15} - 10^{16})$ GeV.

It is easy to see that the superpotential given in equation
(\ref{eq:inflation}),
used to break the rank of the group by one unit, is the same
superpotential used by Dvali et al. \cite{Shafi,Dvali} to implement a false
vacuum hybrid inflationary scenario, identifying the scalar field ${\cal S}$
with the inflaton field. Hence, as shown below,  in supersymmeytric SO(10)
models, the superpotential used to break $U(1)_{B-L}$ can also lead to a
period of inflation. Inflation is then just a consequence of the theory.
In order to understand
the symmetry breaking and the inflationary dynamics, we can study the
scalar potential. The latter is given by (keeping the same notation
for the bosonic component of the superfields as for the superfields):
\begin{equation}
V_1  = |F_S|^2 + |F_{\Phi}|^2 + |F_{\overline{\Phi}}|^2
\end{equation}
where the F-terms are such that $F_{\Psi_i} = |{ \partial W \over \partial
\Psi_i} |$, for $\Psi_i = {\cal S}, \Phi$ and $\overline{\Phi}$. Thus
 \begin{equation}
V_1 = \alpha^2 |{\cal{S}} \overline{\Phi} |^2 + \alpha^2 |{\cal{S}} \Phi |^2 +
|\alpha \overline{\Phi} \Phi - \mu^2|^2 \: .
\end{equation}
The potential is minimized for $arg( \Phi ) + arg(\overline{\Phi}) = 0$,
$(\alpha > 0)$, and it is independent of $\arg({\cal{S}}) + arg({\Phi})$
and $\arg({{\cal{S}}} ) + arg({\overline{\Phi}} )$. From the vanishing
condition of the D-terms, we have $<|\Phi |> = < |\overline{\Phi} |> $. Thus
we can rewrite the scalar potential with the new fields which minimize
the potential, keeping the same notation for the old and new fields,
\begin{equation}
V_1 = 4 \alpha^2 |{\cal{S}}|^2 |\Phi|^2 + (\alpha |\Phi|^2 - \mu^2)^2 \:
\label{eq:V1} .
\end{equation}
The potential has a unique supersymmetric minimum corresponding to
$<|\Phi |> = <| \overline{\Phi} |> = {\mu \over \sqrt{\alpha }}$ and
${\cal S} = 0$. The potential
has also a local minimum corresponding   $S > {\mu \over \sqrt{\alpha}}$ and
$<|\Phi |> =
<| \overline{\Phi} |> = 0$. We identify the scalar field ${\cal S}$ with the
inflaton field and we assume chaotic initial conditions. All the fields are
thus supposed to take initial values the order of the Planck scale, and
 hence the initial value of the inflaton field $ S \gg {\mu \over
\sqrt{\alpha}}$. Since the potential is flat in the ${\cal S}$ direction,
we can minimize it at a fixed value of ${\cal S}$. The $\Phi$ and
${\overline{\Phi}}$ fields roll down their local minimum corresponding
to $<|\Phi |> = <| \overline{\Phi} |> = 0$. The
vacuum energy density is then dominated by a non-vanishing $F_S$-term,
$|F_S| = \mu^2$. The inflationary epoch takes place as the inflaton field
slowly rolls down the potential. $F_S \neq 0$ implies that supersymmetry
is broken. Quantum corrections to the effective potential will help the
fields to slowly roll down their global minimum \cite{Shafi}. At the end
of inflation, the phase transition mediated by the $\Phi$ and
${\overline{\Phi}}$ fields takes place.

Now, in order to break SO(10) down to the standard model gauge group,
we need more than a  $16 + \overline{16}$ or a $126 + \overline{126}$
pair of Higgs fields. We need Higgs in other representations, like the 45, 54
or 210-dimensional
representations if the $Z_2$ parity is to be kept unbroken down to low energy,
as required from proton lifetime measurements. Thus the full superpotential
needed to break SO(10) down to
the standard model must, apart of equation (\ref{eq:inflation}), contains
terms involving the other Higgs needed to implement the symmetry
breaking. Due to the non renormalization theorem in
supersymmetric theories, we can write down the full superpotential
which can implement the desired symmetry breaking pattern, just adding to
equation (\ref{eq:inflation}) terms mixing the other Higgs needed to
implement the symmetry breaking pattern. There can be no mixing between
the latter Higgs and the pair of Higgs used to break $U(1)_{B-L}$ (see
section \ref{sec-building} for example) and the superpotential can be
written as follows :
\begin{equation}
W = W_1({\cal S}, \Phi, \overline{\Phi} ) + W_2(H_1, H_2, ..)
\end{equation}
where ${\cal S}$ is a scalar field singlet under SO(10) identified with
the inflaton field, the $\Phi$ and $\overline{\Phi}$ fields are the
Higgs fields used to break $U(1)_{B-L}$ and the $H_i$ fields, i =
1, .. ,m ,  are the m other Higgs fields needed to implement the full
symmetry breaking pattern  from SO(10) down to the standard model
gauge group. $W_1$ is given by equation (\ref{eq:inflation}) and
$W_1 + W_2$ has a global supersymmetric  minimum such that the
SO(10) symmetry group is broken down to the standard model gauge
group. The scalar potential is then given by
\begin{equation}
V = V_1({\cal S},\Phi, \overline{\Phi}) + V_2(H_i) \: .
\end{equation}
  $V_1$ is given by equation (\ref{eq:V1}) and $V_1 + V_2$ has a
global minimum such that the SO(10) symmetry is broken down to the
standard model gauge group.
 The evolution of the fields
is then as follows. The fields take random initial values, just
subject to the constraint that the energy density is at the Planck
scale. The inflaton field is distinguished from the other fields
from the fact that the GUT potential is
flat in its direction; the potential can be minimized for fixed
${\cal S}$.  Chaotic initial conditions imply that the initial value
of the inflaton field is greater than $\mu \over \sqrt{\alpha}$.
Therefore, the non inflaton fields will roll very
quickly down to their global (or local) minimum, at approximatively a fixed
value for the inflaton, $< |H_i| > \neq 0$, for i = 1,..,n, $< |H_j| > =
0$, for j = n+1,..,m,  and $<|\Phi |> = <|\overline{\Phi} |> = 0$; a
first symmetry breaking,  implemented by the n Higgs fields H acquiring
 VEV, takes place, SO(10) breaks down to an intermeditae symmetry group G.
Then inflation occurs as the inflaton rolls slowly down the potential.
The symmetry breaking implemented with the $\Phi + \overline{\Phi}$
fields occurs at the end of inflation, and the the imtermediate
symmetry group G breaks down to the standard model gauge group.

In the scenario described above, the rank of the intermediate symmetry
group G is equal to the rank of SO(10), which is 5, and hence involves
an unbroken $U(1)_{B-L}$ symmetry. If the rank of the intermediatre
symmetry group were that of the standard model gauge group, that is
if $U(1)_{B-L}$ were broken at the first stage of the symmetry breaking,
the inflationary scenario would unable to solve the monopole problem,
since the later would form at the end of inflation or once inflation
completed. Finally, in models where supersymmetric  SO(10) is broken
directly down to the standard model gauge group, such hybrid inflationary
scenarios cannot cure the monopole problem.

We conclude that if inflation has to occur during the evolution of
the universe
described by a spontaneous symmetric breaking pattern from the
supersymmetric grand unified gauge group SO(10) down to the
minimal supersymmetric standard model, it can thus just be a
consequence of the theory. No external field and no external
symmetry has to be imposed. One can use the superpotential
given in equation (\ref{eq:inflation}) to lower the rank of the
group by one unit and then identify the scalar field ${\cal S}$,
singlet under SO(10), with the inflaton field. A false vacuum
hybrid inflationary scenario will be implemented. It emerges from the theory.

\section{The supersymmetric SO(10) model and the standard cosmology}
\label{sec-topo}

We now construct a supersymmetric SO(10) model which aims to agree with
observations. SO(10) is broken
down to the standard model gauge group with unbroken matter parity
$3_c 2_L 1_Y Z_2$, via the intermediate symmetry group $3_c 2_L 1_R 1_{B-L}$.
We study the symmetry breaking pattern of the model and deduce general impacts
of the model on observations. We look for topological defects formation.

The model initially assumes that the symmetries between particles, forces
and particles, are described by a supersymmetric SO(10) theory. The
SO(10) symmetry is then broken down to the standard model gauge group
via $3_c 2_L 1_R 1_{B-L}$,
\begin{equation}
SO(10) \stackrel{M_{GUT}}{\rightarrow} 3_c 2_L 1_R 1_{B-L}
 \stackrel{M_G}{\rightarrow} 3_c 2_L 1_Y Z_2\stackrel{M_Z}{\rightarrow}
3_c 1_Q Z_2 \: ,
\label{eq:sym}
\end{equation}
$M_{GUT} \sim 10^{16}$ GeV , $M_G \sim M_{GUT}$
with $M_G < M_{GUT}$ and $M_Z \simeq 100$ GeV, and supersymmetry is broken at
$M_s \simeq 10^3$ GeV. The
$Z_2$ symmetry, which appears at the second stage of the symmetry
breaking in (\ref{eq:sym}),  is the discrete$\{1, -1\}$ symmetry,
subgroup of both the $Z_4$ center of $SO(10)$ and of
$U(1)_{B-L}$ subgroup of SO(10). Recal that when we write SO(10) we
really mean its universal covering group spin(10). The $Z_2$ symmetry
acts as matter parity. It preserves large values for the proton
lifetime and stabilizes the LSP; it is
thus necessary that this $Z_2$ symmetry be kept unbroken down to
low energies.

We can look for topological defect formation in the symmetry breaking pattern
using topological arguments \cite{Kibble}. Since Spin(10), the
universal covering group of SO(10), is simply
connected, the second homotopy group
$\pi_2({Spin(10) \over 3_c 2_L 1_R 1_{B-L} }) = \pi_1(3_c 2_L 1_R
1_{B-L}) = Z \oplus Z$ is non-trivial.
Therefore, according to the Kibble mechanism \cite{Kibble}, topological
monopoles form during the first phase transition in equation ({\ref{eq:sym})
when SO(10) breaks down to $3_c 2_L 1_R 1_{B-L}$. They
have a mass $M_m 10^{17}$ GeV. Furthermore, the second
homotopy groups
$\pi_2({Spin(10) \over 3_c 2_L 1_Y Z_2}) = \pi_1(3_c 2_L 1_Y Z_2) = Z$
and  $\pi_2({Spin(10) \over 3_c 1_Q Z_2}) = \pi_1( 3_c 1_Q Z_2) = Z$ are non
trivial so that
the   monopoles are topologically stable down to low energies. These
 monopoles if present today would dominate the energy density of the universe,
and are thus in conflict with cosmological observations.

Now the first homotopy group $\pi_1({3_c 2_L 1_R 1_{B-L} \over 3_c 2_L 1_Q
Z_2})$ is non trivial
and therefore topological cosmic
strings form according to the Kibble
mechanism during the second phase transition in equation({\ref{eq:sym}),
when the $3_c 2_L 1_R 1_{B-L}$ symmetry group breaks down to $3_c 2_L
1_Y Z_2$.  The strings connect half of the monopole-antimonopole pairs
formed earlier \cite{paper}.
Some closed strings can also form. The strings can break with
monopole-antimonopole pair nucleation. The monopoles get attracted
to each other and the whole system of strings disappear
\cite{mon}. Nevertheless,
the other half of the monopoles remain topologically stable.
If present today, these monopoles would lead to a cosmological catastrophe.

Now the rank of $3_c 2_L 1_R 1_{B-L}$ is equal to five, as the rank of SO(10),
and is therefore greater than the rank of
$3_c 2_L 1_Y Z_2$ from one unit. Thus we can couple the inflaton field with the
Higgs
field mediating the breaking of $3_c 2_L 1_R 1_{B-L}$ down to
$3_c 2_L 1_Y Z_2$, see section \ref{sec-inflation}, and the monopole
problem can be cured. If the monopoles are pushed away before the
 phase transition leading to the strings formation takes place,
then the evolution of the string network is quite different than
previously said. It is that of topologically stable cosmic strings.

\section{Model building}
\label{sec-building}
\subsection{Ingredients}

In this section, we explain how to implement the symmetry breaking pattern
given in equation ({\ref{eq:sym}). The model solves the doublet-triplet
splitting and includes an
inflationary scenario as described in section \ref{sec-inflation}.

In order to implement the symmetry breaking pattern given in equation
(\ref{eq:sym}).
and in order to preserve the $Z_2$ symmetry unbroken down to low energy, see
equation (\ref{eq:sym}),  we must only use Higgs fields in ``safe''
representations \cite{Martin}, such as the
adjoint 45, the 54, the 126 or the 210-dimensional representations.

In order to implement the first stage of the symmetry breaking, we could use
only one
 Higgs in the 210-dimensional representation; unfortunately the model
would then not solve the doublet-triplet splitting problem. The latter
can be easily solved using the Dimopoulos-Wilczek mechanism \cite{DW},
using two Higgs, one in the adjoint 45-dimensional representation and
one in the 54-dimensional one. The VEV of the adjoint 45,
which we call $A_{45}$, which implements the Dimopoulos-Wilczek mechanism
is in the {B-L} direction, and breaks SO(10) down to $3_c 2_L 2_R 1_{B-L}$.
The Higgs in the 54 dimensional representation, which we call $S_{54}$, breaks
SO(10) down to $4_c 2_L 2_R$. Altogether the SO(10) symmetry is
broken down to $3_c 2_L 2_R 1_{B-L}$.

We want to break SO(10) directly
down to $3_c 2_L 1_R 1_{B-L}$, we therefore need more
Higgs. We use another 54, which we call $S'_{54}$, and another 45, which
we call $A'_{45}$, in the $T_{3R}$ direction. The latter breaks SO(10) down to
$4_c 2_L 1_R$.  $S'_{54}$ and $A'_{45}$
break together SO(10) down to $4_c 2_L 1_R$.

The role of $S_{54}$ and $S'_{54}$ is to
force $A_{45}$ and $A'_{45}$ into $B-L$ and  $T_{3R}$ directions. SO(10)
breaks down to
$3_c 2_L 1_R 1_{B-L}$ with $A_{45}$, $S_{54}$, $A'_{45}$ ans $S'_{54}$
acquiring VEVs, and as mentioned in section \ref{sec-topo}, topologically
stable monopoles form.

During the second stage of symmetry breaking, see equation (\ref{eq:sym}),
the rank of the group is lowered by one unit. Indeed
the rank of $3_c 2_L 1_R 1_{B-L}$ is equal to the rank of $SO(10)$
which is 5 whereas the rank of $3_c 2_L 1_{Y} Z_2$ is 4. We can
therefore implement a false vacuum hybrid inflationary scenario as
described in section \ref{sec-inflation}, if we couple
the inflaton field to the Higgs field used to break the intermediate
symmetry gauge group $3_c 2_L 1_R 1_{B-L}$. The monopole problem can
be solved and cosmic  strings can form at the end of inflation when
the $3_c 2_L 1_R 1_{B-L}$ symmetry group breaks down to the standard
model gauge group with unbroken matter parity, $3_c 2_L 1_Y Z_2$.

To break $3_c 2_L 1_R 1_{B-L}$, we use a $126 + \overline{126}$ pair
of Higgs fields, which we call $\Phi_{126}$ and $\overline{\Phi}_{126}$.
The latter are safe representations \cite{Martin} and therefore
keeps the $Z_2$ symmetry unbroken. A $16 + \overline{16}$ pair of
Higgs fields usually used for the same purpose would break the $Z_2$
symmetry.  The VEV of the 126 and $\overline{126}$ are in the X direction,
the U(1) symmetry of SO(10) which commutes with SU(5). They break
SO(10) down to $SU(5) \times
Z_2$. All together, i.e. with $A_{45}$, $S_{54}$, $A'_{45}$, $S'_{54}$,
$\Phi_{126}$ and $\overline{\Phi}_{126}$ acquiring VEVs, the SO(10)
symmetry group is broken down to $3_c 2_L 1_Y Z_2$.

The symmetry breaking of the standard model is then achieved using
two Higgs in the $10$-dimensional representation of SO(10), $H_{10}$ and
$H'_{10}$.

To summarize, the symmetry breaking is implemented as follows,
\begin{equation}
SO(10) \stackrel{<A_{45}> <S_{54}><A'_{45}> <S'_{54}>} {\rightarrow}3_c 2_L 1_R
1_{B-L}
\stackrel{<\Phi_{126}> <\overline{\phi_{126}}>} {\rightarrow} 3_c 2_L 1_Y Z_2
\stackrel{<H'_{10}> <H'_{10}>} {\rightarrow} 3_c 1_Q Z_2 \: .
\end{equation}

\subsection{the superpotential}
\label{sec-super}

We now write down the superpotential involving the above mentioned
fields. A consequence of the superpotential is the symmetry breaking pattern
given in
equation (\ref{eq:sym}), which involves an inflationary sector.

As discussed above, our model involves four sectors.
The first sector implements the doublet-triplet splitting and involves
$A_{45}$,  with VEV in the $U(1)_{B-L}$ direction. It also involves
$S_{54}$ and two Higgs 10-plets, $H$ and $H'$.
The superpotential in the first sector is given by $W_1 +W_2$, with,
dropping the subscripts,
\begin{equation}
W_1 = m_A A^2 + m_S S^2 + \lambda_S S^3 + \lambda_A A^2 S
\end{equation}
and
\begin{equation}
W_2 = H A H' + m_{H'} {H'}^2 \: .
\end{equation}
$W_1$ has a global minimum such that the $SO(10)$ symmetry group is
broken down to $3_c 2_L 2_R 1_{B-L}$, with $A_{45}$ and $S_{54}$
acquiring VEVs. $W_2$ implements the doublet-triplet splitting;
$H$ and $H'$ break
$SU(2)_L \times U(1)_Y$ down to $U(1)_Q$.

The second sector involves $A'_{45}$, with VEV in the $T_{3R}$ direction,
and $S'_{54}$.
The superpotential in the second sector is given by,
\begin{equation}
W_3 = m_{A'} {A'}^2 + m_{S'} {S'}^2 + \lambda_{S'} {S'}^3 +
\lambda_{A'} {A'}^2 {S'} \; .
\end{equation}
$W_3$ has a global minimum such that the SO(10) symmetry group is broken down
to
 $3_c 2_L 2_R 1_{B-L}$, with $A'_{45}$ and $S'_{54}$ acquiring VEVs.

The superpotential $W_1 + W_2 + W_3$ has a global minimum such that the SO(10)
symmetry is broken down to $3_c 2_L 1_R 1_{B-L}$, with $A_{45}$, $S_{54}$,
$A'_{45}$ and $S'_{54}$ acquiring VEVs.

The third sector
involves $\Phi_{126}$ and $\overline{\Phi}_{126}$, and breaks $SO(10)$ down
to $SU(5) \times Z_2$. In order to force the $\Phi_{126}$ and
$\overline{\Phi}_{126}$ fields to get their VEVs the order of the GUT scale, we
use a scalar field ${\cal S}$ singlet under SO(10). The superpotential is of
the form, dropping the subscript,
\begin{equation}
W_4 = \alpha {\cal{S}} \overline{\Phi} \Phi - \mu^2 {\cal{S}} \; .
\end{equation}
$\alpha$ and $\mu$ are both positive and we must have
${\mu \over \sqrt{\alpha}} = M_G$, with $M_G \simeq 10^{15} - 10^{16} \: GeV$
for the unification of the gauge coupling constants. Identifying the scalar
field ${\cal S}$ with the inflaton field, $W_4$ leads to a false vacuum hybrid
inflationary scenario, as described in section \ref{sec-inflation}.

The superpotential $W_1 + W_2 + W_3 + W_4$ has a global minimum such that the
$3_c 2_L 1_R 1_{B-L}$ symmetry group is broken
down to the standard model gauge group with unbroken matter parity,
$3_c 2_L 1_Y Z_2$, with $A_{45}$, $S_{54}$, $A'_{45}$, $S'_{54}$,
$\Phi_{126}$ and $\overline{\Phi}_{126}$ acquiring VEVs.

The full superpotential $W = W_1 + W_2 +W_3 + W_4$ does not involve terms
mixing $A'_{45}$ and $S_{54}$, $S'_{54}$ and $A_{45}$ etc... In other
words the three sectors are
independent. Thanks to the non-renormalizable theorem, we are not obliged
to write down these terms, and it is not compatible with any extra
discrete symmetry \cite{babuandbarr}, therefore we do not have to fear
any domain wall formation when the symmetry breaks. Nevertheless, in
order to avoid any undesirable massless Goldstone Bosons, the three
sectors have to be related. The latter can be done introducing a third
adjoint 45 $A"_{45}$, and adding a term of the form $Tr(A A' A'')$ to the
superpotential \cite{babuandbarr}. The latter does neither affect the
symmetry breaking pattern, nor the inflationary scenario discussed
below. The full superpotential of the model is therefore,
\begin{eqnarray}
W &=& m_A A^2 + m_S S^2 + \lambda_S S^3 + \lambda_A A^2 S + H A H' +
m_{H'} {H'}^2 \nonumber \\
&& + m_{A'} {A'}^2 + m_{S'} {S'}^2 + \lambda_{S'} {S'}^3 +
\lambda_{A'} {A'}^2 {S'} \nonumber \\
&& + \alpha {\cal{S}} \overline{\Phi} \Phi - \mu^2 {\cal{S}} + Tr(A A' A'')
\label{eq:super} \: .
\end{eqnarray}
 It leads to the desired
pattern of symmetry breaking and the VEVs of $A_{45}$, $S_{54}$, $A'_{45}$,
$S'_{54}$, $\Phi_{126}$ and $\overline{\Phi}_{126}$ are given as follows
(see appendix). The adjoint $<A_{45}>$ is in th$B-L$ direction,
\begin{equation}
<A_{45}> = \eta \otimes diag(a,a,a,0,0) \label{eq:aaa}
\end{equation}
where $
\eta = \left (
\begin{array} {cc}
0 & 1 \\
-1 & 0
\end{array}
\right )
$ and $a \sim M_{GUT}$. $<S_{54}>$ is a traceless symmetric tensor given by,
\begin{equation}
<S_{54}> = I \otimes diag(x,x,x,-{3\over 2} x,-{3\over 2} x) \label{eq:sss}
\end{equation}
where I is the unitary 2 x 2 matrix and  $x = - {m_A
\over 2 \lambda_A}$.
$<A'_{45}>$ is in the $T_{3R}$ direction,
\begin{equation}
<A'_{45}> = \eta \otimes diag(0,0,0,a',a') . \label{eq:ap}
\end{equation}
where $a' \sim M_{GUT}$. $S'_{54}$ is a traceless antisymmetric tensor,
\begin{equation}
<S'_{54}> = I \otimes diag(x,'x',x',-{3\over 2} x',-{3\over 2} x')
\label{eq:spp}
\end{equation}
where $x' = {2 m_{A'} \over 3 \lambda_{A'}}$.
\begin{equation}
<\Phi_{126}>_{\nu^c \nu^c} = <\overline{\Phi}_{126}>_{\overline{\nu^c}
\overline{\nu^c}} = d \: . \label{eq:ppp}
\end{equation}

With the VEVs chosen above, if $S = 0$, $d = {\mu \over \sqrt{\alpha}}$,
and the superpotential has a global minimum such that the
$SO(10)$ symmetry is broken down to the standard model gauge group with
unbroken amtter parity $SU(3)_c \times SU(2)_L \times U(1)_Y
\times Z_2$, and supersymmetry is unbroken (see the appendix).

\section{The inflationary epoch}
\label{sec-infl}

In this section we evaluate the details of the symmetry breaking pattern
and of the inflationary scenario. We write down the scalar potential and
find values of the scalar coupling constant and the mass scales $M_G$ and
$M_{GUT}$
for which the inflationary scenario is successful.

We are interested in the dynamics of the symmetry breaking pattern and how
the inflationary scenario fits in the symmetry breaking pattern.  We
therefore need to study the scalar potential. In order fully to understand
the dynamics of the model, one would need to use finite temperature field
theory.
Nevertheless, study of the scalar potential derived from the superpotential
given
in equation (\ref{eq:super}) leads a good understanding of the field evolution.
The scalar potential is given by
\begin{eqnarray}
V &=& |2 m_A A + 2 \lambda_A A S|^2 + |2 m_S S + 3 \lambda_S S^2 +
\lambda_A A^2|^2 \nonumber \\
&&+ | 2 m_{A'} A' + 2 \lambda_{A'} {A'} {S'}|^2 + | 2 m_{S'} {S'} + 3
\lambda_{S'} {S'}^2 + \lambda_{A'} {A'}^2 |^2 \nonumber \\
&& +  \alpha^2 |{\cal{S}} \overline{\Phi} |^2 + \alpha^2 | {\cal{S}} \Phi |^2 +
|\alpha \overline{\Phi} \Phi - \mu^2|^2.
\end{eqnarray}
We remind the reader that $A$ and $A'$ are two Higgs in the 45-dimensional
representation of $SO(10)$ with VEV in the $B-L$ and $T_{3R}$ directions
respectively.
$S$ and $S'$ are two Higgs in the 54-dimensional representation
of SO(10). $\Phi$ and $\overline{\Phi}$ are two Higgs in the 126 and
$\overline{126}$ representations, with VEVs in the right-handed neutrino
direction. The scalar field ${\cal S}$
is a scalar field singlet under SO(10). It forces $\Phi$ and $\overline{\Phi}$
to get VEV of the
order of the GUT scale. The scalar field ${\cal S}$ is identified with the
inflaton field.
$\alpha$ and $\mu$ are both positive constants which must satisfy the relation
${\mu \over \sqrt{\alpha}} = M_G$. The potential is minimized for $arg( \Phi )
+ arg(\overline{\Phi}) = 0$,
$(\alpha > 0)$, and it is independent of $\arg({\cal{S}}) + arg({\Phi})$
and $\arg({{\cal{S}}} ) + arg({\overline{\Phi}} )$. We rewrite the
potential with the new fields which minimize the potential, keeping
the same notation for the old and new fields.  The scalar potential becomes
\begin{eqnarray}
V &=& |2 m_A A + 2 \lambda_A A S|^2 + |2 m_S S + 3 \lambda_S S^2 +
\lambda_A A^2|^2 \nonumber \\
&&+ | 2 m_{A'} A' + 2 \lambda_{A'} {A'} {S'}|^2 + | 2 m_{S'} {S'} + 3
\lambda_{S'} {S'}^2 + \lambda_{A'} {A'}^2 |^2 \nonumber \\
&& + 4 \alpha^2 |{\cal{S}}|^2 |\Phi|^2 + (\alpha |\Phi|^2 - \mu^2)^2 +
{1 \over 2} m^2 |{\cal{S}}|^2 \label{eq:scalarpot}
\end{eqnarray}
where we have also introduced a soft
supersymmetry breaking term for $S$, and $m \sim 10^3$ GeV.

The scalar potential is  flat in the
${\cal{S}}$ direction; we thus identify the scalar field ${\cal{S}}$ with the
inflaton field. We suppose chaotic initial conditions; that is
we suppose that all the fields have initial values of the order of the
Planck scale. We then minimize the superpotential for fixed
${\cal{S}}$. We easily find that for ${\cal{S}} > {\mu \over
\sqrt{\alpha}} = s_c$, (recall that $\mu ,\alpha > 0$), there is a local
minimum corresponding to  $|\Phi| = |\overline{\Phi}| = 0$, and $A$, $A'$,
$S$ and $S'$ taking values as
given above in equations (\ref{eq:aaa}), (\ref{eq:sss}), (\ref{eq:ap}) and
(\ref{eq:spp}).
Since all the fields are
assumed to take initial values of the order of the Planck scale, the inflaton
field has an initial value greater than ${\mu \over \sqrt{\alpha}}$.
Then, because the potential is flat in the inflaton direction, the fields
$\Phi$,
$\overline{\Phi}$, $A$, $A'$, $S$ and $S'$ settle quickly to the local
minimum corresponding to $<S>$, $<A>$, $<S'>$ and $<A'>$ as in equations
(\ref{eq:aaa}), (\ref{eq:sss}), (\ref{eq:ap}) and
(\ref{eq:spp}) respectively, and $<|\Phi|> = <|\overline{\Phi}|> =
0$. The first phase transition takes place and the SO(10) symmetry group breaks
down to the $3_c 2_L 1_R 1_{B-L}$ symmetry group. As
shown in section \ref{sec-topo}, topologically stable monopoles form
according to the Kibble mechanism \cite{Kibble} during this first phase
transition.

Once the fields $A$, $S$, $A'$ and $S'$ have settled down to their minimum,
since the first derivatives $\partial V \over \partial A$, $\partial V \over
\partial S$,
$\partial V \over \partial A'$ and $\partial V \over \partial S'$ are
independent of
$\Phi$ and ${\cal{S}}$, the fields $A$, $A'$, $S$, and $S'$ will stay in
their minimum independently of what the fields
$\Phi$ and ${\cal{S}}$ do.
When the VEV of the inflaton field is greater than ${\mu \over
\sqrt{\alpha}} = s_c$, $|\Phi| = |\overline{\Phi}| = 0$, $F_{{\cal{S}}}$
term has a non-vanishing VEV, which means that
supersymmetry is broken in the ${\cal{S}}$ direction, by an amount
measured by the VEV of the ${\cal{S}}$ superfield. There is a
non-vanishing vacuum energy density, $V = \mu^4$. An
inflationary epoch (an exponentially extending universe) can start.

As has been pointed out
recently \cite{Shafi}, the fact that supersymmetry is broken for
$< {\cal S} > \: > \: < {\cal S} >_c$ implies that the one loop
corrections to
the effective potential are non-vanishing. They are given by \cite{Shafi},
\begin{equation}
\Delta V ({\cal{S}}) = \Sigma_i {(-1)^F \over 64 \pi^2} \,
M_i({\cal{S}})^4 \, \ln{({M_i({\cal{S}}) \over \Lambda^2})} \label{eq:effect}
\end{equation}
where the summation is over all helicity states for both fermions and bosons.
$\Lambda$ denotes a
renormalization mass and $(-1)^F$ indicates that the bosons and
fermions make opposite sign contributions to the sum; $(-1)$ stand
for the fermions. Therefore the
one loop effective potential obtained from equations (\ref{eq:scalarpot})
 and (\ref{eq:effect}) is given by \cite{Shafi},
\begin{eqnarray}
V_{eff} &=& \mu^4 [ 1 + {\alpha^2 \over 32 \pi^2} [2 \ln{({\alpha^2
s^2 \over \Lambda^2})} + ({\alpha s^2 \over \mu^2} - 1)^2 \ln{( 1 - {\mu^2
\over \alpha s^2})} \nonumber \\
&& + ({\alpha s^2 \over \mu^2 } + 1)^2 \ln{( 1 + {\mu^2 \over \alpha s^2})} ]
+ {m^2 \over 2 \mu^4} s^2 ] \label{eq:veff}
\end{eqnarray}
where $s = <{\cal S}>$. Now $m \simeq 10^3 \: GeV$ and ${\mu \over
\sqrt{\alpha}} \sim 10^{15 - 16} GeV$, hence unless $\alpha \ll 1$,
the soft supersymmetry breaking term can be neglected. Its contribution
to the scalar potential is negligible. For $s > s_c$, the quantum corrections
to the effective potential help ${\cal{S}}$ to roll down its minimum.
Below $s_c$, the ${\cal{S}}$ field
is driven to zero by the positive
term $\alpha^2 |{\cal{S}}|^2 |\Phi|^2$ which becomes larger
with increasing $|\Phi|$. Rapidly the  $\Phi$, $\overline{\Phi}$ and
${\cal S}$ fields settle down the global minimum of the potential,
corresponding to $<\Phi>_{\nu^c \nu^c} =
<\overline{\Phi}>_{\overline{\nu^c} \overline{\nu^c}} = {\mu
\over \sqrt{\alpha}}$ and $s = 0$. This does not affect the VEVs of the S, A,
S' and A' fields which remain unchanged. The $3_c 2_L 1_R 1_{B-L}$
symmetry group breaks down to $3_c 2_L 1_Y Z_2$. As shown in section
\ref{sec-topo}, topological cosmic
strings form during this phase transition. If inflation ends after the
phase transition, the strings may be inflated away.

Inflation ends when the ``slow roll" condition is violated. The slow
roll condition is characterized by \cite{Ed},
\begin{equation}
\epsilon \ll 1 \; , \;  \eta \ll 1
\end{equation}
where
\begin{equation}
\epsilon = {M_p^2 \over 16 \pi} {({V' \over V})^2} \: ,\qquad  \: \eta =
{M_p^2
\over 8 \pi} {({V'' \over V})} \label{eq:epseta}
\end{equation}
and prime refers to derivatives with respect to s. As pointed out by
Copeland et al. \cite{Ed}, the slow-roll condition is a poor
approximation. But as shown in \cite{Ed}, the number of e-foldings
which occur between the time when $\eta$ and $\epsilon$ reach unity
and the actual end of inflation is a tiny fraction of unity. It is
therefore sensible to identify the end of inflation with $\epsilon$ and
$\eta$ becoming of order unity.

{}From the effective potential (\ref{eq:veff}) and the slow-roll parameters
(\ref{eq:epseta}) we have \cite{Shafi},
\begin{eqnarray}
\epsilon = \left ({\alpha^2 M_p \over 8 \pi^2 M_G}\right )^2 {x^2 \over 16
\pi}
((x^2-1) \, \ln{(1-{1\over x^2})} + (x^2 +1) \ln{(1 + {1\over x})} )^2 \\
\eta = \left ({\alpha M_p \over 2 \pi M_G}\right )^2 {1 \over 16 \pi}
((3 x^2-1) \,
\ln{(1-{1\over x^2})} + (3 x^2 +1) \ln{(1 + {1\over x})} )^2
\end{eqnarray}
where $x$ is such that $s = x s_c$. The phase transition down to the
standard model occurs when $x =1$. The results are as follows. We find
the values of the scalar coupling $\alpha$, the scale $M_{GUT}$
and the scale $M_G$ which lead to successful inflation. For $\alpha
\geq 35 - 43$, $M_G \sim 10^{15} - 10^{16} \: GeV$, $\epsilon$ is always
greater than unity, and the slow roll condition is never satisfied.
The scale $M_{GUT}$ at which the monopoles form satisfies $M_{pl}
\geq M_{GUT} \geq 10^{16} - 10^{17} \: GeV$. For $\alpha \leq 0.02 -
0.002$ and $M_G \sim 10^{15} - 10^{16} \: GeV$, neither $\eta$ nor
$\epsilon$ ever reaches unity. ${\cal S}$ reaches ${\cal S}_c$
during inflation. Inflation must end by the instability of
the $\Phi$ and $\overline{\Phi}$ fields. In that case, inflation
ends in less than a Hubble time \cite{Ed} once ${\cal S}$ reaches
${\cal S}_c$. Cosmic strings, which form when $x=1$,
are not inflated away. The scale $M_{GUT}$ at which the monopoles
form must satisfy $M_{pl} \geq M_{GUT} \geq 10^{16} - 10^{17} \: GeV$.
For the intermediates values of $\alpha$, inflation occurs, and ends
when either $\epsilon$ or $\eta$ reaches unity; the string forming
phase transition takes places once inflation completed.

\section{Formation of cosmic strings}
\label{sec-strings}

In this section we give general properties of the strings which form at the end
of inflation when the $3_c 2_L 1_R 1_{B-L}$ symmetry group breaks down to $3_c
2_L 1_Y Z_2$. We find their
width and their mass, give a general approach for their interactions
with fermions and study their superconductivity.

\subsection{General properties}

Recall that, since the  first homotopy group
$\pi_1({3_c 2_L 1_R 1_{B-L} \over 3_c 2_L 1_Y Z_2})$ is non trivial, cosmic
strings form during the second phase transition (see equation
\ref{eq:sym}) when the $3_c 2_L 1_R 1_{B - L}$ symmetry breaks down to
$3_c 2_L 1_Y Z_2$ .
We note that the subspace
spanned by $R$ and $B - L$ is also spanned by $X$ and $Y$. The generator
of the string corresponds to the U(1) of SO(10) which commutes with
SU(5), and the gauge field forming the string is the corresponding
gauge field, which we call X. The strings are abelian and physically
viable. The model does not give rise to
Alice strings, like most of the non abelian GUT phase transitions
where abelian and non abelian strings form at the same time. This is a
good point of the model, since Alice strings give rise to quantum
number non conservation, and are therefore in conflict with the
standard cosmology. The strings arising in our model can be related to the
abelian strings arising in the symmetry breaking pattern of SO(10)
down to to the standard model with $SU(5) \times Z_2$ as intermediate
scale, since they have the same generator; the latter have been widely
studied in the non-supersymmetric case \cite{Aryal,abelian}. Nevertheless,
in our model, inside the core of the string, we do not have an SO(10)
symmetry restoration, but an $3_c 2_L 1_R
1_{B - L}$ symmetry restoration. We therefore don't have any gauge fields
mediating baryon number violation inside the core of the strings, but one
of the fields violates $B-L$. We also expect the supersymmetric strings to
have different properties and different interaction with matter due
to the supersymmetry restoration inside the core of the string.
These special properties will be studied elsewhere.

The two main characteristics of the strings, their width and their mass,
are determined through the Compton wavelength of the Higgs and gauge
bosons forming the strings. The Compton wavelength of the Higgs and gauge
bosons are
respectively
\begin{equation}
\delta_{\Phi_{126}} \sim m_{\Phi_{126}}^{-1} = ( 2 \: \alpha \:  M_G )^{-1}
\label{eq:deltaphi}
\end{equation}
and
\begin{equation}
\delta_X \sim m_X^{-1} = (\sqrt{2} \: e \: M_G)^{-1} \label{eq:deltaX}
\end{equation}
where e is the gauge coupling constant in supersymmetric $SO(10)$ and
it is given by ${e^2 \over 4 \pi} = {1 \over 25}$ and $M_G$ is the scale at
which the strings form.

As mentioned above, the strings formed in our model can be related
to those formed during the symmetry breaking pattern $SO(10)
\rightarrow SU(5) \times U(1) \rightarrow SU(5) \times Z_2$. These
strings have been studied by Aryal and Everett \cite{Aryal} in
the non supersymmetric case. Using their results, with appropriate
changes in the gauge coupling constant and in parameters of the Higgs
potential, we find that the string mass per unit length of the string
is given by :
\begin{equation}
\mu \simeq (2.5-3) \: (M_G)^2 \label{eq:mu} \: ,
\end{equation}
for the scalar coupling $\alpha$  ranging from $5 \times 10^{-2}$ to
$ 2 \times 10^{-1}$. Recall that the mass per unit- ength characterizes
the entire properties of a network of cosmic strings.

\subsection{No Superconducting Strings}

One of the most interesting feature of GUT strings is their
superconductivity. Indeed, if they become superconducting at the
GUT scale, then vortons can form and dominate the energy density of the
universe; the model loses all its interest! The strings arising in
our model are not superconducting in Witten's sense
\cite{Witten}. They nevertheless can become current carrying with
spontaneous current generation at the electroweak scale through
Peter's mechanism \cite{Patrick}. But it is believed that this does
not have any
disastrous impact on the standard cosmology. It has been shown in the
non-supersymmetric case that the abelian strings arising when SO(10)
breaks down to $SU(5) \times Z_2$ have right-handed neutrino zero
modes \cite{Das}. Since the Higgs field forming the string is a Higgs
in the 126 representation which gives mass to the right-handed
neutrino and winds around the string, we expect the same zero modes on
our strings. Since supersymmetry is restored in the core of the
string, we also expect bosonic zero modes of the superpartner of the
right-handed neutrino. Now, the question of whether or not the string
will be current carrying will depend on the presence of a primordial
magnetic field, and the quantum charges of the right-handed neutrino
with respect to this magnetic field. If there is a primordial magnetic
field under which the right-handed neutrino has a non-vanishing
charge, then the current will be able to charge up. On the other hand,
if such magnetic field does not exist, or if the right-handed neutrino is
neutral, then there will be nothing to generate
the current of the string.
Although it is possible to produce a primordial magnetic field in
a phase transition \cite{Tanmay}, we do not expect the fields produced
 through the mechanism of reference \cite{Tanmay} to be able to charge up
 the current  on the string, since the latter are correlated on too large
scales. Nevertheless, the aim of this section is to show that the strings will
not
be superconducting at the GUT scale in any case. We can therefore
assume a worse situation,
that is, suppose that the magnetic fields are correlated on smaller
scales,
due to any mechanism for primordial magnetic field production any time
after the Planck scale. In our model, cosmic strings form when $3_c
2_L 1_R 1_{B - L}$ breaks down to $3_c 2_L 1_Y Z_2$. Therefore the symmetric
phase  $3_c 2_L 1_Y Z_2$ will be associated with
 color, weak and hypercharge magnetic fields.  The color and weak
magnetic fields formed when SO(10) broke down to $3_c 2_L 1_R 1_{B - L}$,
and and the hypercharge
magnetic field formed at the following phase transition, formed from
the R and $B-L$ magnetic fields.
Since the charges of the right-handed neutrino with respect to the color,
weak and hypercharge magnetic fields are all vanishing, no current will
be generated.

We conclude that the strings will not be superconducting at the GUT scale.
They might become superconducting at the electroweak scale, but this does
not seem to affect the standard big-bang cosmology in any essential way.

If the strings formed at the end of inflation are still present today, they
would affect temperature fluctuations in the CBR and have affected large scale
structure formation.

\section{Observational consequences}
\label{sec-observations}

We show here that the strings formed at the end of inflation may be present
today.
We find the scale $M_G$ at which cosmic strings form and
the scalar coupling of the inflaton field which are consistent with
the temperature fluctuations observed by COBE. We then examine the
dark-matter content of the model and make a qualitative discussion
regarding large scale structure formation.

\subsection{Temperature fluctuations in the CBR}
\label{sec-temp}
If both inflation and cosmic strings are part of the scenario,
temperature fluctuations in the CBR are the result of the quadratic
sum of the temperature fluctuations from inflationary perturbations
and cosmic strings.

The scalar density perturbations produced by the inflationary epoch
induce temperature fluctuations in the CBR which are given by \cite{Shafi},
\begin{eqnarray}
\left ({\delta T \over T} \right )_{inf} & \simeq & { \sqrt{ 32 \pi \over
45}} {
V^{3 \over 2}
\over V' M_{pl}^3 } |_{x_q} \\
& \approx & (8 \pi N_q )^{1 \over 2} \left ({M_G \over M_{pl}} \right )^2
\label{eq:DTinfl}
\end{eqnarray}
where the subscript indicates the value of ${\cal S}$ as the scale
(which evolved to the present horizon size) crossed outside the
Hubble horizon during inflation, and $N_q$ ($\sim 50 - 60$) denotes
the appropriate number of e-foldings. The contribution to the CBR
anisotropy due
to gravitational waves produced by inflation in this
model is negligible.

The cosmic strings density perturbations also induce CBR anisotropies given
by \cite{Allen},
\begin{equation}
\left ({\delta T \over T} \right )_{c.s.} \approx 9 \: G \mu \label{eq:DTcs}
\end{equation}
where $\mu$ is the strings mass per unit length, which is given by
equation (\ref{eq:mu}). It depends on the scalar coupling $\alpha$. Since the
later is undetermined, we can use the order of magnitude
\begin{equation}
\mu \sim \eta^2 \label{eq:app}
\end{equation}
which holds for a wide range of the parameter $\alpha$, see equation
(\ref{eq:mu}) and in ref. \cite{Aryal}. In eq. (\ref{eq:app}), $\eta$ is the
symmetry breaking scale associated with the strings
formation, here $\eta = M_G$.

Hence, from equations (\ref{eq:DTinfl}) and (\ref{eq:DTcs}) the temperature
fluctuations in the CBR are given by,
\begin{eqnarray}
\left ({\delta T \over T} \right )_{tot} &\approx &  \sqrt{\left
({\delta T \over
T}\right )_{inf}^2 + \left ({\delta T \over T}\right )_{c.s.}^2 }
\label{eq:DT1}\\
 &\approx & \sqrt{ 8 \pi N_q + 81} \:
\left ({M_G \over M_{pl}} \right )^2 \label{eq:DT}
\end{eqnarray}
The temperature fluctuations from both inflation and cosmic strings
add quadratically. Since they are both proportional to ${M_G \over
M_{pl} }$ their computation is quite easy.

An estimate of the coupling
$\alpha$ is obtained from the relation \cite{Shafi},
\begin{equation}
{\alpha \over x_q} \sim { 8 \pi^{3\over 2} \over \sqrt{N_q}} {M_G \over
M_{pl}} . \label{eq:alpha}
\end{equation}

With $x_q \sim 10$, using equations (\ref{eq:DT}) and (\ref{eq:alpha}) and
using the density fluctuations measured by COBE $\simeq 1.13 \times
10^{(-5)}$ \cite{Bennet} we get
\begin{eqnarray}
\alpha & \simeq & 0.03 \\
M_G & \simeq & 6.7 \: 10^{15} \: GeV \: .
\end{eqnarray}
With these values, we find that $\eta$ reaches unity when $x \simeq 1.4$
and the scale $M_{GUT}$ at which the monopoles form must satisfy,
\begin{equation}
M_{pl} \geq M_{GUT} \geq 6.7 \: 10^{16} \: GeV
\end{equation}
where $M_{pl}$ is the Planck mass $\simeq 1.22 \: 10^{19} \: GeV$.

{}From the above results, we can be confident that the strings forming
at the end of inflation should still be around today.

Now that we have got values for the scalar coupling $\alpha$ and the
scale $M_G$ at which the strings form, the Compton wavelength of the
Higgs and gauge bosons forming the strings given by equations
(\ref{eq:deltaphi}) and (\ref{eq:deltaX}) can be computed. We find
\begin{equation}
\delta_{\Phi_{126}} \sim m_{\Phi_{126}}^{-1} \sim 0.42 \: 10^{-28} \: cm
\end{equation}
for the Compton wavelength of the Higgs field forming the string and
\begin{equation}
\delta_X \sim m_X^{-1} \sim 0.29  \: 10^{-29} \: cm
\end{equation}
for the Compton wavelength of the strings gauge boson.
$\delta_{\Phi_{126}} > \delta_X$ thus the strings possess an inner
core of false vacuum of radius $\delta_{\Phi_{126}}$ and a magnetic
flux tube with a smaller radius $\delta_X$. The string energy per unit
length is given by equation (\ref{eq:mu}),
thus, using above results, we have
\begin{equation}
G \mu \sim 7.7 \: 10^{-7}
\end{equation}
where $G$ is Newton's constant. The results are slightly affected by the
number of e-folding and by the order of magnitude (\ref{eq:app}) used to
compute the temperature fluctuations in the CBR due to cosmic strings
in equations (\ref{eq:DT1}) and (\ref{eq:DT}). Once we have found the value
for the scalar coupling $\alpha$ for successful inflation, we can redo
the calculations with a better initial value for the string mass per
unit length, see equation (\ref{eq:mu}); the scalar coupling $\alpha$
is unchanged. The results are summarized in Table 1.

{\bf Table 1} : The table shows the values obtained for the scale $M_G$
at which the strings form, the scalar coupling $\alpha$, the Higgs and
gauge boson Compton wavelengths $\delta_\Phi$ and $\delta_X$ of the
strings, and $G \mu$, where $\mu$ is the strings mass per unit
length and $G$ is the Newton's constant,
 for different values of the number of e-foldings $N_q$ and for different
initial values used for their computation for the string
mas-per-unit length.

\vspace{.2cm}

\begin{tabular} {|c|c|c|c|c|}
\hline
Nq & 50 & 50 & 60 & 60 \\ \hline
$\mu_{init}$ & $\eta^2$ & $2.5 \: \eta^2$ & $\eta^2$ & $2.5 \: \eta^2$ \\
\hline
$M_G$ & $6.8 \: 10^{15}$ & $6.3 \: 10^{15}$ & $6.5 \: 10^{15}$ & $6.1 \:
10^{15}$  \\ \hline
 $\alpha$ & 0.03 &  0.03 & 0.03 & 0.29 \\ \hline
$\delta_\Phi$ & $0.42 \: 10^{-28}$ & $0.44 \: 10^{-28}$ &$0.43 \: 10^{-28}$
&$0.46 \: 10^{-28}$ \\ \hline
$\delta_X$ & $0.29 \: 10^{-29}$ & $0.31 \: 10^{-29}$ &$0.30 \: 10^{-29}$ &$0.32
\: 10^{-29}$ \\ \hline
$G \mu$  & $7.7 \: 10^{-7}$ &$6.7 \: 10^{-7}$ & $7.1 \: 10^{-7}$ & $6.3 \:
10^{-7}$ \\ \hline
\end{tabular}

\subsection{Dark Matter}
\label{sec-dark}

We specify here the nature of dark matter generated by the model.

If we go back to
the symmetry breaking pattern of the model given by equation
(\ref{eq:sym}), we see that a discrete $Z_2$ symmetry remains
unbroken down to low energy. This $Z_2$ symmetry is a subgroup of both
the $Z_4$ center of SO(10) and of $U(1)_{B-L}$ subgroup of SO(10).
This $Z_2$ symmetry acts as matter parity. It preserves large values
for the proton lifetime and stabilizes the Lightest Super Particle.
The LSP is a good Cold Dark Matter candidate.

The second stage
of symmetry breaking in equation (\ref{eq:sym}) is implemented with the
use of a $126 + \overline{126}$ pair of Higgs multiplets, with VEVs in the
direction of the $U(1)_X$ of $SO(10)$ which commutes with $SU(5)$. The
$\overline{126}$-multiplet can couple with fermions and give
superheavy Majorana mass to the right-handed neutrino, solving the solar
neutrino problem via the MSW mechanism \cite{MSW} and providing a good Hot Dark
Matter candidate. This can be done if all fermions are assigned to
the 16-dimensional spinorial representation of SO(10). In that case, couplings
of the form $f \overline{\Psi} \Psi \overline{126}$,
where $\Psi$ denotes a 16-dimensional spinor to which all fermions belonging to
a single family are assigned, provide right-handed
neutrinos masses of order $m_{R} \simeq  10^{12}$ GeV,
if $f \sim 10^{-4}$ GeV. Neutrinos also get Dirac masses which are typically of
the order of the mass of the up-type quark of the corresponding family; for
instance $m^{\nu^e}_D \simeq m_u$. After diagonalizing the neutrino mass
matrix, one finds that the right-handed neutrino mass $m_{\nu_R} \simeq m_R$
and the left-handed neutrino mass $m_{\nu_L} \simeq {m_D^2 \over m_R}$. With
the above values we get
\begin{eqnarray}
m_{\nu^e} &\sim &  10^{-7} \: eV\\
m_{\nu^\mu} &\sim & 10^{-3} \: eV\\
m_{\nu^\tau} &\sim & 10 \: eV
\end{eqnarray}
The tau neutrino is a good Hot Dark Matter candidate.

Our model thus provides both CDM and HDM and is consistent with mixed Cold and
Hot DM scenarios.

It is interesting to note that CDM and HDM are, in this model, related to each
other. Indeed, the $Z_2$ symmetry in equation (\ref{eq:sym}), which stabilizes
the LSP, is kept unbroken because a $126 + \overline{126}$ and not  $16 +
\overline{16}$ pair of Higgs fields are used to break $U(1)_{B-L}$. If a 16
dimensional Higgs representation were used,
the right-handed neutrino could not get a superheavy Majorana mass
and thus no HDM could be provided, also the $Z_2$ symmetry would
have been broken, and thus the LSP destabilized. The $126 +
\overline{126}$ pair of Higgs fields provide superheavy Majorana
to the right-handed neutrino and keeps the $Z_2$-parity unbroken.
It leads to both HDM and CDM. We conclude that, in this model,
CDM and HDM are
intimately related. Either the model provides both Cold and Hot
Dark Matter, or it does not provide any. Our model provides
both CDM and HDM.

\subsection{Large Scale Structure}
\label{sec-structure}

We give here only a qualitative discussion of the consistency of the model
with large scale structure. We do not make any calculations which would
require a full study on their own. We can nevertheless use various results
on large scale structure with inflation or cosmic strings. Since
we determined the nature of dark matter provided by the model, we may make
sensible estimations about the the consistency of the model with large
scale structure.

Presently there are two candidates for large scale structure formation, the
inflationary scenario and the topological defects scenario with cosmic
strings. Both scenarios are always considered separately. Indeed,
due to the difference in the nature of the density perturbations in
each of the models, density perturbation calculations due to a mixed
strings and inflation scenario are not straightforward. Indeed in the
inflation-based models density perturbations are Gaussian adiabatic
whereas in models based on topological defects inhomogeneities are
created in an initially homogeneous background \cite{Andrew}.

In the attempt to explain large scale structure, inflation-seeded
Cold Dark Matter models or strings models with HDM are the most
capable \cite{Andrew}. In adiabatic perturbations with Hot Dark
Matter small scale perturbations are erased by free streaming
whereas seeds like cosmic strings survive free streaming and
therefore smaller scale fluctuations in models with seeds + HDM
are not erased, but their growth is only delayed by free streaming.

Our model involves both Hot and Cold Dark Matter, and both inflation
and cosmic strings. It is therefore sensible to suggest that our model
will be consistent with large scale structure formation, with the large
scale fluctuations resulting from the inflationary scenario and
small scale fluctuations being due to cosmic strings.

\section{Conclusions}
\label{sec-conclusion}

We have successfully implemented a false vacuum hybrid inflationary
scenario in a supersymmetric SO(10) model. We first argued that this
type of inflationary scenario is a natural way for inflation to occur in
global supersymmetric SO(10) models. It is natural, in
the sense that the inflaton field emerges naturally from the theory, no
external field and no external symmetry has to be added. The scenario does not
require any fine
tuning. In our specific model, the SO(10) symmetry is broken via the
intermediate $3_c 2_L 1_R
1_{B-L}$ symmetry down to the standard model with unbroken matter parity $3_c
2_L 1_Y Z_2$. The model
gives a solution for the doublet-triplet splitting via the
Dimopoulos-Wilczek mechanism. It also suppresses rapid proton decay .

The inflaton,
a scalar field singlet under SO(10), couples to the Higgs mediating the
 phase transition associated with the breaking of $3_c 2_l 1_R 1_{B-L}$ down to
the standard model. The scenario starts with chaotic initial conditions.
The SO(10) symmetry breaks at $M_{GUT}$ down to $3_c 2_L
1_R 1_{B-L}$ and topologically stable monopoles form.
There is a non-vanishing vacuum energy density, supersymmetry is broken,
 and an exponentially extending epoch starts. Supersymmetry is broken,
 and therefore quantum corrections to the scalar potential can not be
neglected. The latter help the inflaton field to roll down its minimum.
At the end of inflation the $3_c 2_L 1_R
1_{B-L}$ breaks down to $3_c 2_L 1_Y Z_2$, at a scale $M_G$, and cosmic strings
form. They are not superconducting.

Comparing the CBR temperature anisotropies measured by COBE with that
predicted by the mixed inflation-cosmic strings scenario, we find  values
for the scalar coupling $\alpha$ and for the scale $M_G$ at which the
strings form. $M_{GUT}$ is calculated such that we get enough e-foldings
to push the monopoles beyond the horizon. The results are summarized in Table
1.
 The evolution of the strings is that of topologically stable cosmic
strings.
The model is consistent with a mixed HCDM scenario. Left-handed neutrinos
get very small masses and the tau neutrino may serve as a good HDM candidate.
They could also explain the solar neutrino problem via the MSW mechanism.
The unbroken matter parity stabilizes the LSP, thus providing a good
CDM candidate. A qualitative discussion leads to the conclusion that
the model is consistent with large scale structures, very large scale
structures being explained by inflation and cosmic strings explaining
structures on smaller scales. An algebraic investigation for this purpose would
be useful, but will require further research.

\section*{Acknowledgements}
I am grateful to Q. Shafi for useful discussions and I thank
the Isaac Newton Institute for encouraging discussions/collaborations
between participants. I would also like to thank R.R. Caldwell and
N. Manton. I acknowledge Newnham College and PPARC for financial support.

\appendix
\section{Minimizing the Superpotential}

In this appendix, we find the true minimum of the superpotential of
the model. We calculate the F-terms and find the VEVs of the Higgs
fields which correspond to the global minimum.

The full superpotential of the model is given by equation (\ref{eq:super}),
\begin{eqnarray}
W &=& m_A A^2 + m_S S^2 + \lambda_S S^3 + \lambda_A A^2 S + H A H' +
m_{H'} {H'}^2 \nonumber \\
&& + m_{A'} {A'}^2 + m_{S'} {S'}^2 + \lambda_{S'} {S'}^3 +
\lambda_{A'} {A'}^2 {S'} \nonumber \\
&& + \alpha {\cal{S}} \overline{\Phi} \Phi - \mu^2 {\cal{S}} + Tr(A A' A'') \;
{}.
\end{eqnarray}
where $A$ and $A'$ are 54-dimensional Higgs representations therefore
traceless second rank antisymmetric tensors. Thus in the 10-dimensional
representation of SO(10) they are of the form, with appropriate subscripts,
\begin{equation}
<S_{54}> = I \otimes diag(x,x,x,-{3\over 2} x,-{3\over 2} x) \label{eq:vevs}
\end{equation}
and
\begin{equation}
<S'_{54}> = I \otimes diag(x,'x',x',-{3\over 2} x',-{3\over 2} x')
\label{eq:vevsp}
\end{equation}
where $x$ and $x'$ are the order of $M_{GUT}$ and are determined by the
vanishing condition of the F-terms.
The Higgs $A_{45}$ and $A'_{45}$ are 45-dimensional representations and must be
in
the $B-L$ and $T_{3R}$ directions respectively (see section
\ref{sec-building}). Therefore in the 10-dimensional representation of
SO(10) $A_{45}$ and $A'_{45}$ are given by,
\begin{equation}
<A_{45}> = \eta \otimes diag(a,a,a,0,0) \label{eq:veva}
\end{equation}
where $a \sim M_{GUT}$ and
\begin{equation}
<A'_{45}> = \eta \otimes diag(0,0,0,a',a') . \label{eq:vevap}
\end{equation}
where $a' \sim M_{GUT}$. The $\Phi$ and $\overline{\Phi}$ fields are Higgs in
the $126$ and $\overline{\Phi}$-dimensional representations. The $\Phi$ and
$\overline{\Phi}$ fields must break the $U(1)_X$ symmetry which commutes with
SU(5), and thus acquire VEVs in the right-handed neutrino direction. From the
vanishing condition for the D-terms, $<\Phi > = <\overline\Phi >$ and thus,
with appropriate subscripts,
\begin{equation}
<\Phi_{126}>_{\nu^c \nu^c} = <\overline{\Phi}_{126}>_{\overline{\nu^c}
\overline{\nu^c}} = d \label{eq:vevphi}
\end{equation}
where $d \sim M_G$.

The true vacuum corresponds to F-terms vanishing. Supersymmetry is
unbroken. Using the same notation for the scalar component than for
the superfield, the F-terms are given by,
\begin{eqnarray}
F_A &=& 2 m_A A + 2 \lambda_A A S \\
F_S &=& 2 m_S S + 3 \lambda_S S^2 + \lambda_A A^2 \\
F_{A'} &=& 2 m_{A'} A' + 2 \lambda_{A'} {A'} {S'} \\
F_{S`} &=& 2 m_{S'} {S'} + 3 \lambda_{S'} {S'}^2 + \lambda_{A'} {A'}^2 \\
F_\Phi &=& \alpha {\cal{S}} \overline{\Phi} \\
F_{\overline{\Phi}} &=& \alpha {\cal{S}} \Phi \\
F_{{\cal{S}}} &=& \alpha \overline{\Phi} \Phi - \mu^2
\end{eqnarray}
Using the VEVs of the Higgs given above, we easily get the VEV's of
the F-terms. The vanishing
condition for the latter leads to the following relations, for
each
term respectively,
\begin{eqnarray}
 m_A a + 2 \lambda_A a x &=& 0 \label{eq:une}\\
 - m_S x + {3 \over 4} \lambda_S x^2 + {1 \over 5} \lambda_A a^2 &=& 0
\\
2 m_{A'} a' - 3 \lambda_{A'} a' x' &=& 0 \\
-m_{S'} x' + {3 \over 4} \lambda_{S'} {x'}^2 - \lambda_{A'} {a'}^2 &=&
0 \\
\alpha s d &=& 0 \\
\alpha d^2 - \mu^2 &=& 0 \label{eq:deux}
\end{eqnarray}
where s is the VEV os the scalar field ${\cal S}$. We note that the roles
of the 54 dimensional representations $S_{54}$ and $S'_{54}$ are to force the
adjoint $A_{45}$ and $A'_{45}$ into $B-L$ and $T_{3R}$ directions. With the
VEVs
chosen above, see equations (\ref{eq:vevs}),
(\ref{eq:veva}),  (\ref{eq:vevsp}),  (\ref{eq:vevap}) and
(\ref{eq:vevphi}), if $s=0$ and $d={\mu \over \sqrt{\alpha}}$ the potential
has a global
minimum, such that the SO(10) symmetry is broken down to $SU(3)_c
\times SU(2)_L \times U(1)_R \times U(1)_{B - L}$ and supersymmetry is
unbroken and we have $x = {2 m_{A} \over 3 \lambda_{A}}$ and $x' =
{2 m_{A'} \over 3 \lambda_{A'}}$.  $a \sim M_{GUT}$, $a' \sim M_{GUT}$
and ${\mu \over \sqrt{\alpha}} \sim
M_G$ where $M_G \sim 10^{15 - 16}$ GeV and $M_G \leq M_{GUT}
\leq M_{pl}$
and $M_{pl}$ is the Planck mass $\sim 10^{19}$ GeV.

\end{document}